\documentclass{ws-procs9x6}

\def\etal{{\it et al.}}
\def\ifm#1{\relax\ifmmode#1\else$#1$\fi}
\def\f{\ifm{\phi}}    
\def\ff{\f--factory}
\def\epm{\ifm{e^+e^-}}
\def\x{\ifm{\times}}
\renewcommand{\to}{\ensuremath{\rightarrow}}
\def\order#1,{\ifm{\mathcal{O}(10^{#1})}}
\def\dafne{DA$\Phi$NE}
\def\ep{\ifm{e^+}}
\def\el{\ifm{e^-}}
\def\ks{\ifm{K_S}}
\def\kl{\ifm{K_L}}
\def\ko{\ifm{K_0}}
\def\ok{\ifm{\overline{K}_0}}

\def\pip{\ifm{\pi^+}}
\def\pim{\ifm{\pi^-}}

\def\ab{\ifm{\sim}}
\def\MeV{\ifm{\mbox{MeV}}}
\def\GeV{\ifm{\mbox{GeV}}}
\def\mum{\ifm{\mbox{$\mu$m}}}
\def\mm{\ifm{\mbox{mm}}}
\def\mub{\ifm{\mbox{$\mu$b}}}
\def\invfb{\ifm{\mbox{fb}^{-1}}}
\def\JPC{\ifm{J^{PC}}}
\newcommand{\ket}[1]{\ifm{|#1\rangle}}
\newcommand{\braket}[2]{\ifm{\langle #1|#2 \rangle}}
\def\Dt{\ifm{\Delta t}}
\def\CPT{CPT}

\begin{document}

\title{ STATUS AND PROSPECTS FOR LORENTZ AND CPT VIOLATION
TESTS AT KLOE AND KLOE-2 }

\author{ANTONIO DE SANTIS}

\address{Dipartimento di Fisica, Universit\`a di Roma 
`La Sapienza' \\
and I.N.F.N. Sezione di Roma \\
P.le A. Moro, 2\\ 
I-00185 Rome, Italy \\ 
E-mail: antonio.desantis@roma1.infn.it}

\author{on behalf of the KLOE and KLOE-2 collaborations}

\begin{abstract}
The neutral kaon system offers a unique possibility to perform fundamental 
tests of \CPT\ invariance. In this contribution the KLOE prospects for 
the measurements of \CPT\ violation in the context of the Standard-Model 
Extension are presented together with a full description of the analysis 
method needed and with the perspective given by the KLOE-2 data-taking 
campaign.

\end{abstract}

\bodymatter

\section{The KLOE experiment}\label{ads:sec:intro}

The KLOE experiment operates at \dafne, the 
Frascati \ff. \dafne\ is an \epm\ collider running at 
a center of mass energy of \ab1020 \MeV, the mass of the 
\f\ meson. Positron and electron beams collide at an 
angle of $\pi$-25 mrad, producing \f\ mesons with small boost in 
the orbit plane ($p_x(\f) \ab-15$ MeV).
\par
The KLOE detector consists of a large cylindrical drift chamber (DC) 
surrounded by a lead-scintillating fiber electromagnetic calorimeter (EMC). 
A superconducting coil around the EMC provides a 0.52 T axial field. 
The DC\cite{DCH} is 4 m in diameter and 3.3 m long and has 12,582 
all-stereo tungsten sense wires. The chamber 
shell is made of carbon fiber-epoxy composite and the gas used is a 90\% 
helium, 10\% isobutane mixture. These features maximize transparency to 
photons and reduce \kl\to\ks\ regeneration and multiple scattering. The 
position resolutions are $\sigma_{r,\phi}$\ab 150 \mum\ and $\sigma_z$\ab 2 \mm. 
The momentum resolution is $\sigma(p_{\perp})/p_{\perp}\ab 0.4\%$. 
The calorimeter\cite{EMC} is divided into a barrel and two endcaps, 
and covers 98\% of the solid angle. The modules are read-out at both ends by 
photomultipliers, both in amplitude and time for a total of 2440 cells 
per side arranged in five layers.   
Cells close in time and space are grouped into calorimeter 
clusters. The cluster energy $E$ is the sum of the cell energies. 
The cluster time $T$ and position $\vec{R}$ 
are energy-weighed averages. Energy and time resolutions are $\sigma_E/E = 
5.7\%/\sqrt{E(\GeV)}$ and  $\sigma_t = 57{\rm ps}/\sqrt{E(\GeV)} 
\oplus 100\ {\rm ps}$, respectively.
\par
Actually KLOE has already acquired 2.5 \invfb\ of data and a new 
extensive campaign of data taking is starting aiming at an 
integrated luminosity of 25 \invfb.
\par
The cross section for the process \ep\el\to\f\ production 
is 3.3 \mub\ and the \f\ meson
decays into \ko\ok\ with a branching fraction of \ab34\%.
The initial state of the kaon pair is produced via strong 
interaction with quantum numbers $\JPC=1^{--}$:
\begin{equation} \label{ads:eq:initstate}
  \ket{i} = \frac{\ket{\ko}\ket{\ok}-
    \ket{\ok}\ket{\ko}}{\sqrt{2}}
  = \mathcal{N}(\ket{\ks}\ket{\kl}-
  \ket{\ks}\ket{\kl}),
\end{equation}
where $\ket{\ks/\kl}=(1+\varepsilon_{S/L})\ket{\ko} \pm 
(1-\varepsilon_{S/L})\ket{\ok}$
and $\varepsilon_{S/L}=\varepsilon\pm\delta$.
The two kaons are in an antisymmetric correlated state and 
the time evolution of their state in two different final 
states $f_1$ and $f_2$ for the two kaons as a function of the 
difference of decay time ($\Dt = t_2-t_1$) can be expressed 
as the decay intensity as a function of \Dt:
\begin{eqnarray}
I_{f_1f_2}(\Dt) & \propto & |\eta_1| e^{-\Gamma_L \Dt} + |\eta_2|e^{-\Gamma_S \Dt} 
\nonumber \\ 
& & 
-2 e^{-\frac{(\Gamma_S+\Gamma_L)}{2}\Dt}\cos( \Delta m \Dt + \phi_2-\phi_1),
\label{ads:eq:timeevo} 
\end{eqnarray}
where 
\begin{eqnarray}
\eta_j & = & \braket{f_j}{\kl}/\braket{f_j}{\ks}=|\eta_j|e^{i\phi_j}. \label{ads:eq:decampli} 
%\end{equation}
\end{eqnarray}
\section{CPT and Lorentz symmetry breaking}
A general theoretical possibility for \CPT\ violation is based on spontaneous
breaking of Lorentz symmetry, as developed by 
Kosteleck\'y\cite{kost1,kost2,kost3}: the Standard-Model Extension (SME).
\par
In SME for neutral kaons, \CPT\ manifests to lowest order only in the 
\CPT\ violation parameter $\delta$, and exhibits a dependence on the 4-momentum 
of the kaon:
\begin{equation} \label{ads:eq:deltak}
\delta \approx i \sin \phi_{SW} e^{i \phi_{SW}} \gamma_K (\Delta a_0-
\vec{\beta}_K \cdot\Delta{\vec{a}})/\Delta m,
\end{equation}  
where $\gamma_K$ and $\vec{\beta}_K$ are the kaon boost factor and 
velocity in the observer frame, and $\Delta a_{\mu}$ are the \CPT\ 
violation coefficients for the two valence quarks in the kaon.
\par 
The time dependence arising from the rotation 
of the Earth can be explicitly displayed in \eref{ads:eq:deltak} by 
choosing a three-dimensional basis ($\hat{X},\hat{Y},\hat{Z}$) in a 
nonrotating frame, with the $\hat{Z}$ axis along the Earth's rotation axis, 
and a basis $(\hat{x},\hat{y},\hat{z})$ for the laboratory frame\cite{kost2}.
The \CPT\ violating parameter $\delta$ may then be expressed as:
\begin{eqnarray} 
%\begin{equation} 
%\begin{array}{l}
%\delta(\vec{p}_K,t) = 
%\frac{i \sin\phi_{SW}e^{i\phi_{SW}}}{\Delta m} \\
%\gamma_K\left\{ \Delta a_0
%+ \beta_K { \Delta a_Z}(\cos\vartheta\cos\chi-\sin\vartheta\cos\phi\sin\chi) \right.\\
% -\beta_K \Delta a_X \sin\vartheta\sin\phi\sin\Omega t \\
%+ \beta_K \Delta a_X(\cos\vartheta\sin\chi+\sin\vartheta\cos\phi\sin\chi)\cos\Omega t \\
%+ \beta_K \Delta a_Y (\cos\vartheta\sin\chi+\sin\vartheta\cos\phi\sin\chi)\sin\Omega t \\
% \left. -\beta_K \Delta a_Y\sin\vartheta\sin\phi\cos\Omega t 
%\right\} \\
%\end{array}
%\label{ads:eq:delta_sid}
%\end{equation}
\delta(\vec{p}_K,t) &=& 
\frac{i \sin\phi_{SW}e^{i\phi_{SW}}}{\Delta m} \gamma_K
\Big[ \Delta a_0
+ \beta_K { \Delta a_Z}(\cos\vartheta\cos\chi-\sin\vartheta\cos\phi\sin\chi)
\nonumber \\
&&
\hskip 80pt
 -\beta_K \Delta a_X \sin\vartheta\sin\phi\sin\Omega t 
\nonumber \\
&& 
\hskip 80pt
+ \beta_K \Delta a_X(\cos\vartheta\sin\chi+\sin\vartheta\cos\phi\sin\chi)\cos\Omega t 
\nonumber \\
&&
\hskip 80pt
+ \beta_K \Delta a_Y (\cos\vartheta\sin\chi+\sin\vartheta\cos\phi\sin\chi)\sin\Omega t 
\nonumber \\
&&
\hskip 80pt
-\beta_K \Delta a_Y\sin\vartheta\sin\phi\cos\Omega t 
\Big],
\label{ads:eq:delta_sid}
\end{eqnarray}
where $t$ is the sidereal time, $\Omega$ is the Earth's sidereal 
frequency and $\chi$ is the angle between the Earth rotation axis
and the $\hat{z}$ direction in the laboratory frame.
\par
Since inside the KLOE detector, kaons are produced in almost all direction, 
it is possible to use the experiment like a telescope to explore any 
direction in the space by using two different method to measure the 
$\Delta a_\mu$ parameters of \eref{ads:eq:deltak}:
\begin{itemize}
\item by studying the sidereal time variation of the 
  semileptonic asymmetry of the \kl\ and \ks. 
  A preliminary measurement of the $\Delta a_0$ has been already 
  performed with KLOE data using the difference between the 
  semileptonic asymmetries of \ks\ and \kl: 
  $\Delta a_0 = (0.4\pm1.8)\x10^{-17} \GeV$\cite{CPT07_didoLNF}.
\item by studying the quantum interferometry between the two kaons, 
  \eref{ads:eq:timeevo}. In this case, two different final state 
  can be considered: ($\pi l^+\nu$,~$ \pi l^-\nu$) 
  and ($\pip\pim$,~$\pip\pim$).
\end{itemize}
\par
In the following we will focus on interferometry with the channel 
\f\to\ks\kl\to\pip\pim\pip\pim. Detailed discussion concerning
other methods can be found elsewhere\cite{CPT07_didoLNF}.
Using this channel the two kaons decay both
in the same final state, but have opposite momenta and experience 
different direction in the space. For instance, we can distinguish 
between the two kaons by their forward or backward emission. 
In this way it is possible to define as $K_+$ $(K_-)$ the one going in the 
forward (backward) direction with $cos(\vartheta)>(<)$ $0$.
The ratio of amplitudes in \eref{ads:eq:decampli} is
$\eta_\pm=\varepsilon - \delta(\vec{p})$. Since the kaon momenta
are opposite, a small asymmetry in the decay intensity $I_\pm(\Dt)$ 
will appear and the following ratio will be different from
zero:
\begin{equation}\label{ads:eq:fbasymmetry}
A(\Dt)= \frac{I_\pm(\Dt>0)-I_\pm(\Dt<0)}{I_\pm(\Dt>0)+I_\pm(\Dt<0)}.
\end{equation}
The above asymmetry for $\Dt \gg \tau_S$ tends to zero, 
because $\varepsilon$ and $\delta$ are $90^{\circ}$ out of 
phase\cite{buchanan}:
\begin{equation}\label{ads:eq:fbasymmetry_long}
A(\Delta t \tau_S)\simeq -2\Re\left( \frac{\delta}{\epsilon}\right)\sim 0,
\end{equation}
while for $|\Delta t|\leq 5 \tau_S$
\begin{equation}\label{ads:eq:fbasymmetry_short}
A(| \Delta t | 5\tau_S) \propto -2\Im m\left( \frac{\delta}{\epsilon}\right)
\end{equation}
and therefore to $\Delta a_{X,Y,Z}$.
\par
With analysis based on 1 \invfb\ we get a preliminary 
result on $\Delta a_{X,Y,Z}$\cite{discrete08}:
\begin{eqnarray}
  \Delta a_X = (-6.3\pm 6.0)\times 10^{-18}\;\GeV, \nonumber\\ 
  \Delta a_Y = (2.8\pm 5.9)\times 10^{-18}\;\GeV ,\nonumber\\ 
  \Delta a_Z = (2.4\pm 9.7)\times 10^{-18}\;\GeV. 
\end{eqnarray}
\par
This analysis scheme using the simple forward backward asymmetry is not 
sensitive to $\Delta a_0$ and its effect is washed out in the asymmetry
\eref{ads:eq:delta_sid} and \eref{ads:eq:fbasymmetry}. 
As shown in \eref{ads:eq:deltak} this parameter is coupled only 
with the $\gamma_K$ factor. Since at \dafne\ the \f\ is produced with a 
small boost in the horizontal plane the kaons have different values for
$\gamma_K$ as a function of the azimuthal angle. The ratio  
of the decay intensities $I_\pm(\Dt)$ distribution for events
in which the $K_+$ propagates opposite or along the \f\ momentum
will enhance the small asymmetry introduced by the $\Delta a_0$
component of the $\delta$ parameter. For values of $\Delta a_0$ of
the order \order-18, we expect an effect on the $I_\pm(\Dt)$ up to
1\%-2\% in the region $|\Dt|<5\tau_S$ from which we expect to be able 
to put limits of the order \order-17,-\order-18,. 
\par
The $\Delta a_\mu$ parameters can be all simultaneously measured 
by performing a proper sidereal time  dependent analysis of 
asymmetries of $I_\pm(\Dt)$, \eref{ads:eq:timeevo} and 
\eref{ads:eq:delta_sid}.
An accuracy $\order-18,\;\GeV$ could be reached with the analysis 
of the full KLOE data sample.

\section{Conclusions and future plans}
All four $\Delta a_{\mu}$ parameters of the SME can be independently 
measured at KLOE, completing results obtained by fixed beam experiments. 
\par
The continuation of the KLOE physics program with 
KLOE-2\cite{AmelinoCamelia:2010me} 
at an improved DA$\Phi$NE machine is currently starting. 
The data taking campaign will be organized in two different steps. 
During the first one we will double the statistics already taken by KLOE with 
a new beam interaction scheme\cite{Milardi:2010bm} and with the inclusion of 
two pairs of electron-positron taggers\cite{Babusci:2009sg} for the 
study of the gamma-gamma 
physics. The second phase aims to an integrated luminosity of \ab25 \invfb\
including several upgrades for the KLOE detector: 
\begin{itemize}
\item a pair of crystal calorimeters (CCALT\cite{Happacher:2009xm}) near 
  the interaction 
  region to improve the angular acceptance for low-$\theta$ particles;
\item a pair of tile calorimeters (QCALT\cite{Cordelli:2009xb}) 
  covering the quadrupoles along the beam pipe made of tungsten 
  foil and singly read-out scintillator tiles to improve the angular 
  coverage for particles coming from the active volume of the DC;
\item a small and light tracker (IT\cite{Archilli:2010xb}) made of four planes of cylindrical GEM 
  to improve the resolution of the vertex reconstruction around the 
  interaction point and to increase the low-$\theta$ charged particles
  acceptance.
\end{itemize}
\par
One of the main physics issue of KLOE-2 is the search for \CPT\ violation
effects; limits on several parameters are expected to be improved by about 
one order of magnitude.

\end{document}